\documentclass{article}

\usepackage[english]{babel}

\usepackage[letterpaper,top=2cm,bottom=2cm,left=3cm,right=3cm,marginparwidth=1.75cm]{geometry}

\usepackage{amsmath}
\usepackage{amssymb}
\usepackage{graphicx}
\usepackage[colorlinks=true, allcolors=blue]{hyperref}

\title{Unraveling 20th-century political regime dynamics using the physics of diffusion}
\author{Paula Pirker-Díaz$^{1\ast}$,
	Matthew C. Wilson$^{2}$,
	Sönke Beier$^{1}$,
    Karoline Wiesner$^{1}$\and
	\small$^{1}$Institute of Physics and Astronomy, University of Potsdam, Potsdam, Germany.\and
	\small$^{2}$Department of Political Science, University of South Carolina, Columbia, USA.\and
	\small$^\ast$Corresponding author. Email: paula.pirker.diaz@uni-potsdam.de\and}

\begin{document}
\maketitle

\begin{abstract}
Uncertainty persists over how and why some countries become democratic and others do not, or why some countries remain democratic and others ‘backslide’ toward autocracy. Furthermore, 
while scholars generally agree on the nature of `democracy' and `autocracy', the nature of regimes in-between -- and changes between them -- are much less clear. 
By applying the spectral dimensionality-reduction technique Diffusion Map to political-science data from the V-Dem project for the period 1900 to 2021, we identify a low-dimensional non-linear manifold on which all electoral regimes move. Using the diffusion equation from statistical physics, we measure the time scale on which countries change their degree of electoral quality, freedom of association, and freedom of expression depending on their position on the manifold.
By quantifying the coefficients of the diffusion equation for each country and over time,  we show that democracies behave like sub-diffusive (i.e. slow spreading) particles and that autocracies on the verge of collapse behave like super-diffusive (i.e. fast spreading) particles. 
We show that regimes in-between exhibit diffusion dynamics distinct from autocracies and democracies, and an overall higher instability.
Furthermore, we show that a country's position on the manifold and its dynamics are linked to its propensity for civil conflict. 
Our study pioneers the use of statistical physics in the analysis of political regimes. 
Our results provide a quantitative foundation for developing theories about what changes during democratization and democratic backsliding, as well as a new framework for regime-transformation and risk-of-conflict assessment.
\end{abstract}

\section{Introduction}

The study of democracy and democratization lies at the center of political science and is increasingly important in other social science disciplines. In the post-Cold War world, democracy promotion has also become a central foreign-policy objective for many countries and is often a critical condition for the distribution of international developmental aid. 
% A 
Yet, uncertainty persists over how and why some countries become democratic and others do not \cite{LevitskyWay:2010}, or why some countries remain democratic and others ‘backslide’ \cite{haggard2021backsliding,waldner2018unwelcome}.
% B 
We also lack generalizable theories to explain why some autocratic regimes are remarkably stable over many decades, before suddenly collapsing \cite{mounk:2016} -- the most prominent example being the fall of the Soviet Union in 1989. 
% C 
Almost all existing methods to quantify democracy vs autocracy put them at opposite ends of a uni-dimensional measure. An alternative approach is to treat democracy as a binary concept (e.g., Przeworski et al. 2000\cite{Przeworskietal:2000}). A standard approach to assessing the quality of democracy is to aggregate expert judgments about institutional attributes such as media independence or election capacity into an index. The V-Dem Institute's Electoral Democracy Index \cite{vdemcodebook}, Freedom House' Democracy Score \cite{freedomhouse2024}, and the Polity score \cite{MarshallJaggers:2002} 
are the most common and widely used examples.
Problems persist about the use of such indices, however -- namely, that the nature of regimes ‘in-between’ democracy and autocracy remains unclear. Terms such as ‘electoral autocracy’, ‘illiberal democracy’, or ‘pseudo-democracy’ are regularly used to describe states that are neither completely autocratic nor considered fully democratic, with no agreed-upon terminology or underlying theory about them \cite{hyde2020democracy}. 
 
For a long time, theoretical developments about qualities of democracy were hampered by a data problem. This changed substantially with the start of the Varieties of Democracy (V-Dem) dataset project (\href{https://v-dem.net}{v-dem.net}) . The V-Dem project uses a large number of expert surveys and a Bayesian measurement model to generate quantitative estimates for over 190 countries between the years 1789 and 2023. The project produces estimates about specific institutional attributes such as `election integrity' or `media censorship' \cite{vdemmethods}, providing detailed assessments of different features that are considered crucial for democratic function. 
These attributes, also known as indicators, are aggregated into "high-level" indices such as the Electoral Democracy Index (EDI), which rates from 0 to 1 the level of `democraticness' of each country in a given year. \cite{vdemcodebook, vdemmethods}
In recent work, we showed that the high-dimensional V-Dem data offer more insights than an aggregate one-dimensional index would, by showing that election capability plays a key role in stabilising electoral autocracies \cite{wiesner2024principal}. Though this is well known in the literature on authoritarian regimes, it is largely hidden in composite measures that estimate `democraticness'.

Here, we significantly advance our understanding of democracy and democratization by
% A
observing the dynamics of countries in the 20th-century  on a non-linear manifold that we constructed from the V-Dem data using the so-called \emph{Diffusion Map} (DM) technique \cite{dm-main,coifman2005geometric}.  
% B
Using the anomalous diffusion equation from statistical physics, we show that the dynamics of democracies differ significantly from those of autocracies and in-between regimes. Using the diffusion equation, we quantify the propensities of movement and find a clear distinction for electoral autocracies both in their stability as well as in their amount of change once they begin to open up. In the language of physics, we show that electoral autocracies on the verge of breaking down are similar to super-diffusive particles, while consolidated democracies behave like sub-diffusive particles. 
% C
Our statistical-physics approach allows us to quantitatively distinguish non-democracies from democracies and electoral autocracies according to their position on the manifold and their propensity of movement on the manifold. 

We use a subset of the V-Dem data that relates to electoral democracy, which comprises 25 variables on election quality, suffrage, freedom of association, and freedom of expression on 172 countries between the years of 1900 and 2021 -- 12,296 (country-year) data points in total. A list of the variables and a description of each is given in Tab. \ref{tab:var-correl_psi1} in the \hyperref[appendix:data]{Methods section}.
Our methods rely on the nonlinear dimensionality-reduction technique \textit{Diffusion Map} \cite{dm-main,coifman2005geometric} and the application of the diffusion equation from statistical physics. Using the Diffusion Map, we identify a low-dimensional non-linear manifold in the V-Dem data on which all electoral regimes move. Using the statistical physics of diffusion, we measure the time scale on which countries changed their degree of electoral quality, freedom of association, and freedom of expression depending on their position on the manifold. By doing so, we identify distinct diffusive behaviours depending on the regime type, providing a refined characterisation of the dynamics of political states over time.

Our results offer a solid quantitative foundation on which theories about democratization, democratic backsliding, and extreme political events can be built.  
Not least, the results have important implications for international development. 
We show a link between a country's position on the manifold and its propensity for civic conflict, thus offering a new suggestion for assessing conflict risk. Shifts in a state's dynamics also serve as proxies for regime transformation risks, signaling potential moments of political instability that policymakers can more precisely target given its location on the manifold. Lastly, uncovering the factors that characterise democratic regression offers a more targeted approach to democratic preservation and conflict prevention. Together, these insights offer a new framework for anticipating and mitigating political risks that acknowledges the myriad ways that states can change in a complex system.

\section{Results}
\subsection*{The political diffusion manifold of the 20th century constructed from the V-Dem data}

\begin{figure}[h!]
\centering
\includegraphics[width=.65\linewidth]{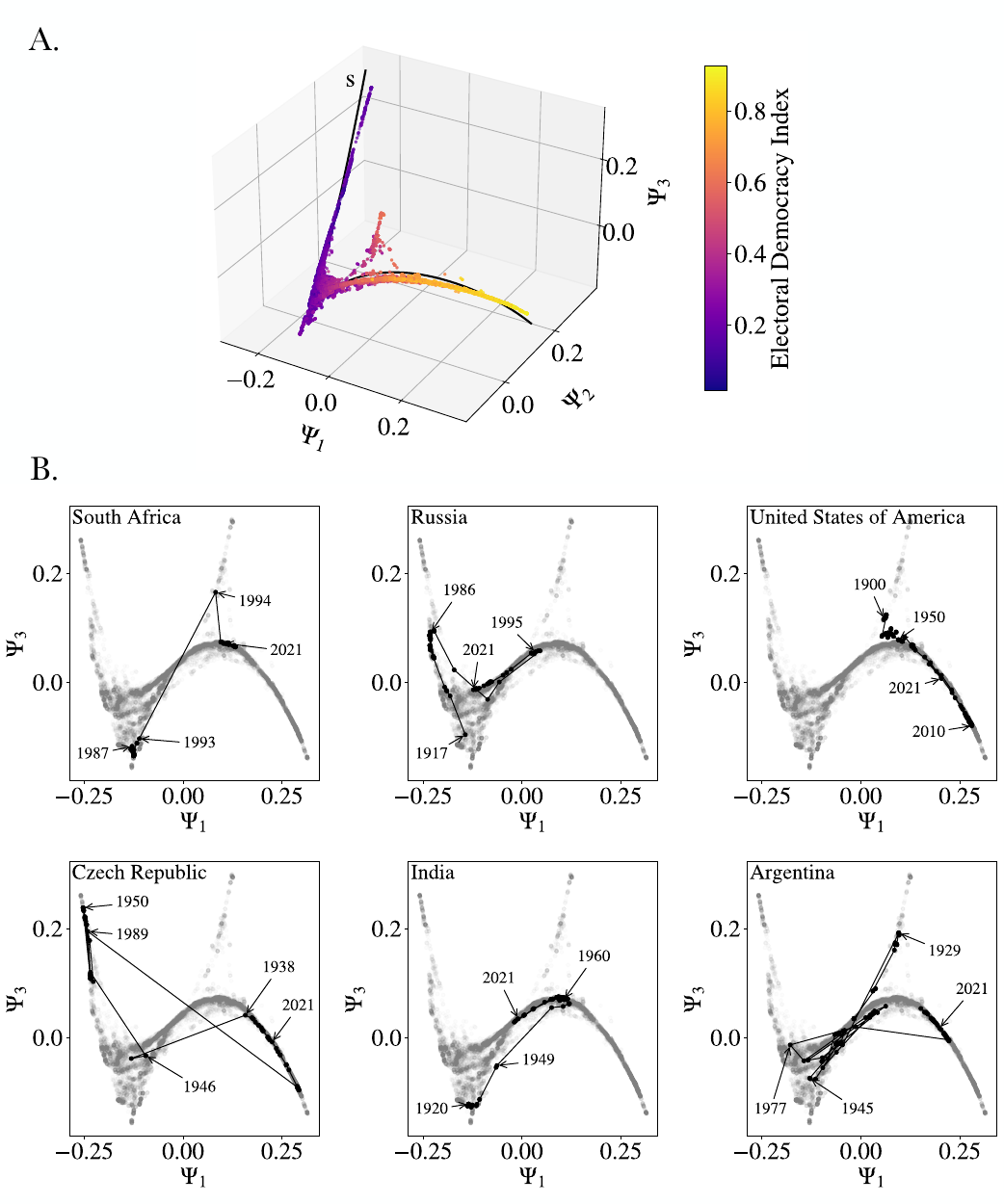}

\caption{A. Diffusion manifold of V-Dem data for 174 countries for the years 1900 to 2021 (projected onto first three components, color coded according to V-Dem's Electoral Democracy Index). Each dot is one country-year data point (12,296 data points in total). In black: a polynomial curve fit  ($s$) to the DM manifold (for details, see the corresponding \hyperref[appendix:fit]{Methods section}).
B. Selected time trajectories on diffusion manifold (projected onto first and third component): South Africa (1910-2021), Russia (1917-2021), United States of America (1900-2021), Czech Republic (1920-2021), India (1920-2021) and Argentina (1904-2021).}
\label{fig:manifold}
\end{figure}

We applied the nonlinear dimensionality-reduction technique \textit{Diffusion Map} (DM) to the 12,296 country-year events of the V-Dem data set, resulting in $n$ components (dimensions), where $n$ is the size of the data set --- see the corresponding \hyperref[appendix:dm]{Methods section} for a more detailed and technical description. The first DM component parametrizes the most elongated direction of the manifold. In other words, for a 1-D manifold, the first component fully parametrizes the entire manifold, and for higher dimensional manifolds further components are needed. \cite{Nadler-dm-chapter}
The resulting manifold is shown in Fig.~\ref{fig:manifold} in the projection onto the first three DM components: $\Psi_1, \Psi_2$ and $\Psi_3$.   
The manifold is almost one-dimensional, with the exception of two triangular structures. 
The country-year events in Fig.~\ref{fig:manifold} are color-coded according to their EDI value, indicating that more fully developed democracies are located at one end of the manifold, while countries further from democratic ideals are located at the other end. $\Psi_1$ effectively serves as a proxy. This is confirmed by the Spearman and Pearson correlation coefficients between EDI and $\Psi_1$, which are 0.9629 and 0.9699, respectively.

Political states evolve over time, which we can visualize by tracing each country's position on the manifold across the available years. 
In Fig.~\ref{fig:manifold} (bottom figures), we show the time trajectory of six selected countries that occupied very different positions on the manifold: South Africa, Russia, the U.S., Czech Republic, India and Argentina. 
Note that the DM technique is ignorant of time (it does not know that 1967 comes after 1966), and yet there is apparent order associated with historical developments in the data. The U.S. exhibits a gradual, diffusion-like evolution. This example gradually became more democratic between 1900 and the 2010s, after which regression occurred. 
Others undergo dramatic shifts, with large jumps in short periods of time, exemplified by Argentina and the Czech Republic. The major jumps in Argentina's trajectory are attributable to military coups that ousted civilian regimes; in case of the Czech Republic, the dramatic shifts correspond to transitions to and away from one-party communist rule (between 1948 and 1989). Other countries, such as South Africa, experience a mix of both gradual movement and large jumps, depending on the years considered. South Africa changed dramatically at the end of its Apartheid regime, which is clearly visible in the large jump on the manifold between 1993 and 1994. 
In Russia and India, we see evidence of both liberalisation and regression, with different jump sizes in different regions of the manifold (see Fig. \ref{fig:manifold}).  Russia -- then the Soviet Union -- developed into a strong one-party regime after 1922 until the regime's collapse in 1991, although the country has since become more autocratic. Likewise, India substantially progressed following independence in 1947 but has also experienced a regression as a result of increasing ruling-party dominance.

The country examples in Fig.~\ref{fig:manifold}
highlight the diverse pathways that countries take through the manifold. Based on common conceptualisations of regimes that occurred over these countries' political histories, we ascertain that electoral autocracies -- defined by minimal suffrage and the use of elections by a ruling party to maintain power -- occupy one extreme end of the manifold and that open democracies exist at the other. Low-capacity and occupied states, by contrast, lie in a `trench' between the two. The examples also show that the extent of countries' movements across the manifold may be influenced by their position on it. 
This suggests that the manifold not only characterises distinct political regimes but also reveals dynamic patterns in political evolution. The DM analysis indicates that regime type and dynamic type are related.

\begin{figure}[h!]
\centering
\includegraphics[width=.9\linewidth]{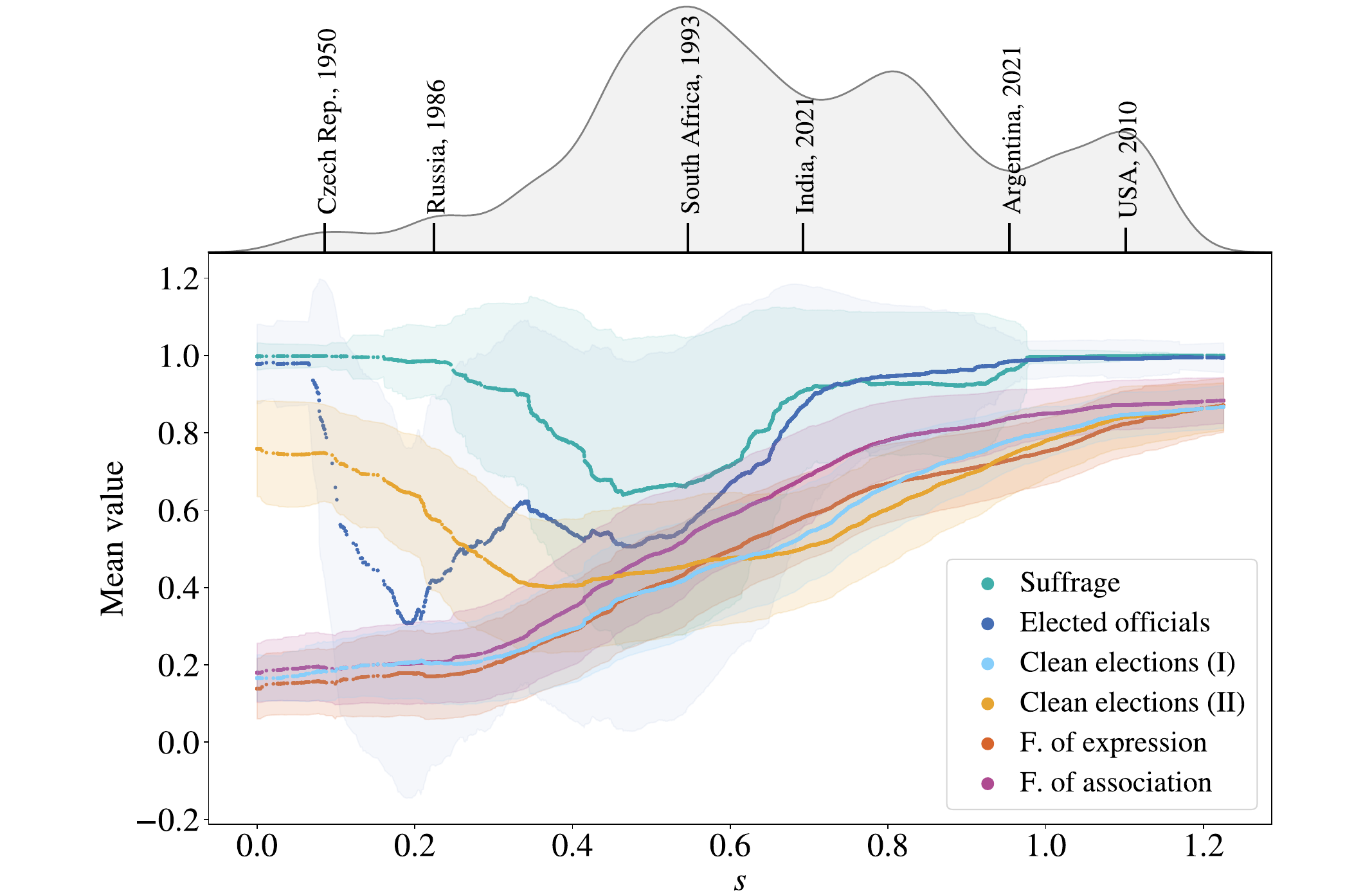}
\caption{Mean and averaged standard deviation of the V-Dem variables of each group along the diffusion manifold (parametrised through $s$), computed using a sliding window of width $\Delta s = 0.2$. Variables are grouped into: clean elections (I) and (II), freedom of expression, freedom of association, elected officials and  suffrage (see text and Tab. \ref{tab:var-correl_psi1} for details). A Kernel Density Estimate of the distribution of data points (i.e. country-years) along $s$ and labels indicating six selected country-year examples are shown on top.}
\label{fig:evolutions} 
\end{figure}

Due to the non-linearity of the DM-technique, the relative contributions of the V-Dem variables to the EDI vary along the manifold. To unpack these contributions, 
we fitted a polynomial curve in  three-dimensional space (see detailed expression in the corresponding \hyperref[appendix:fit]{Methods section} and visualisation in Fig.~\ref{fig:manifold}, A). Due to the manifold's quasi-one-dimensional structure,  we obtained an approximation  of the manifold in terms of a single parameter which we call $s$. 
For each of the 25 V-Dem variables, we computed the mean and standard deviation along the `unfolded' manifold, as parametrized by $s$, using a sliding window of width $\Delta s=0.2$. Next, following V-Dem's use of lower-level indices to represent distinct concept groups within the Electoral Democracy Index, we separate the 25 variables into six groups: clean elections (separated into I and II), freedom of expression, freedom of association, suffrage and elected officials (the `elected officials' and the `suffrage' variable are both lower-level indices, see V-Dem documentation \cite{vdemcodebook} and the corresponding \hyperref[appendix:data]{Methods section}.. These groups correspond to V-Dem's lower-level indices, with the exception of the groups `clean elections' I and II (which together form V-Dem's lower-level `clean elections' index) \cite{vdemcodebook}.
For a list of variables in each group, see the \textit{Indicator classification group} column in Tab.~\ref{tab:var-correl_psi1} of the \hyperref[appendix:data]{Methods section}. 
Fig.~\ref{fig:evolutions} shows the mean value and the averaged standard deviation for each group as a function of $s$. 
As noted earlier, very low values of $s$ correspond to electoral autocracies and very high values of $s$ to democracies. This is illustrated at the top of Fig.~\ref{fig:evolutions} with select years from our country examples embedded in a kernel-density estimate of the country-year distribution along $s$. 
The figure reveals that not all attributes contribute equally to developments from autocracy to democracy (and vice versa) -- instead, different groups of V-Dem variables contribute differently along the DM manifold. 

For $s>0.6$, all groups exhibit an increasing trend, reflecting the expected correlation between higher democratic quality and higher values in all attributes. However, being positioned at the other end of the manifold (extreme left) does not necessarily imply low values for all variables. In fact, `suffrage', `elected officials' and `clean elections' (II) show high mean values in this region, where electoral autocracies are located. This confirms the idea that election outcomes are completely controlled by governments in electoral autocracies (countries placed in the low-$s$ region). Our results are in line with the principal-component analysis by Wiesner et al. \cite{wiesner2024principal}, but give a much more detailed picture of the roles that elections, elected officials, and suffrage play (or don't play) in the democratic quality of a regime. Moving further along the manifold, into 
the range $s\in (0.2, 0.5)$, there is a decreasing trend and greater variability in the groups `suffrage', `elected officials' and `clean elections (II)', while the groups `freedom of expression', `freedom of association', and `clean elections (I)' begin to show an increased slope in their continuous upward trajectory. This is evidence for the loss of control of the government over the election outcomes. Notably, the variables in the `clean elections (I)' group pertain to the extent to which the regime disadvantages opposition parties, either by inhibiting free and fair elections, through intimidation, or by controlling the autonomy of election management.  The `clean elections (II)' group, by contrast, more closely represents the capacity of the regime to effectively carry out an election. Moving along the manifold, the autonomy of the electoral management body increases and the government intimidation improves (aspects captured in `clean elections (I)'), but this loss of control results in an increase of voting irregularities, violence and vote buying, among others (which is represented by a decrease in `clean elections (II)'). This change occurs alongside improvements in aspects related to the freedoms of expression and association, which enhances civil society. For this reason, the ruling party or leader in transitioning regimes may attempt to limit citizens' ability to decide the outcome, resulting in diminished suffrage.

The evolution of the averaged standard deviation also depends on the group of variables and the region of $s$ we are looking at. Two relevant cases are `suffrage' and `elected officials', which show large standard deviations. `Suffrage' shows higher standard deviation in the middle region, where countries with restricted suffrage are located. In contrast, `elected officials' shows a large standard deviation along the whole spectrum with the exception of the right extreme (full democracies). This reflects the fact that the countries in the middle of the manifold are a diverse lot in terms of citizen inclusion in elections. Likewise, authoritarian regimes that rely on elections to govern notably do not allow all offices to be competitive -- instead, different countries allow citizen input on select positions, such as a portion (or all) of the legislature.

\subsection*{Anomalous diffusion approach}

Our choice of the DM as methodology is not merely a convenience but bears deeper meaning. In the following section, we show that the changes of political regimes in the 20th century follow the mathematical laws of anomalous diffusion. 
Anomalous diffusion is a generalization of regular (Brownian) motion, often occurring in systems where standard diffusion laws (like Fick’s law) don’t fully capture the particle dynamics. The classic diffusion equation, based on Fick's law, assumes a \emph{linear relationship} between the mean squared displacement (MSD) of particles and time: 
$
\langle x^2(t) \rangle \simeq K t
$.

In anomalous diffusion, however, this relationship is modified to a \emph{power law}:

\begin{equation}\label{eqn:diffeq}
    \langle x^2(t) \rangle \simeq K_\alpha t^{\alpha}
\end{equation}

where:
\begin{itemize}
    \item $ \alpha = 1 $ indicates normal diffusion,
    \item $ \alpha < 1 $ signifies \emph{sub-diffusion} (particles move slower, often due to obstacles or traps),
    \item $ \alpha > 1 $ signifies \emph{super-diffusion} (particles spread faster, as in systems with long-range correlations or Lévy flights). 
\end{itemize}
The constant $K_\alpha$, the so-called generalized diffusion coefficient, defines the scale of the movement, being larger for larger average step sizes. 
Diffusion processes are observed across various systems, from atoms in magneto-optical traps to the movement of bacteria or animals foraging. \cite{Muñoz-Gil:2021, Vilk:2022} 

In the following we will group particles locally in the diffusion space and compute the MSD in the 25-dimensional original space. We proceed as follows.

\begin{figure}[ht]
\centering
\includegraphics[width=.97\linewidth]{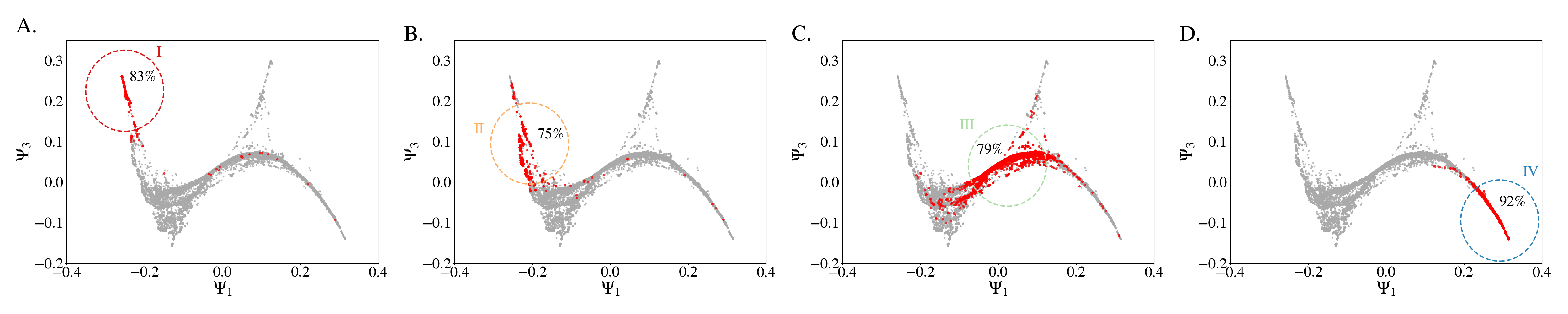}
\caption{Movement within three years of all data points initially in spherical neighborhood of radius $\rho=0.1$ around one of three selected country-years: (A) Albania 1946 (case I), (B) Rwanda 1989 (case II), (C) Argentina 1961 (case III) and (D) Norway 2000 (case IV). The percentage of countries initially within the neighbourhood and still remaining after three years is shown for each case.}
\label{fig:diffusion1}
\end{figure}

For each data point (i.e., country-year) we examine the dynamics within its neighborhood, defined by a sphere of radius $\rho=0.1$ in the 3-dimensional space formed by $\Psi_1$, $\Psi_2$, and $\Psi_3$. Fig. \ref{fig:diffusion1} shows the neighbourhoods of four selected data points in the $\Psi_1$ and $\Psi_3$ projection. Neighboring points in the diffusion map are \emph{similar} in the original 25-dimensional space, making this a natural grouping for analyzing dynamics within comparable regimes. 
Furthermore, in Fig. \ref{fig:diffusion1} we consider all country-year events within the given neighborhoods and display their position three years later (see markers in red). This visualization emphasizes the connection between a country’s position in the manifold and its likelihood of undergoing a political state change. 

\begin{figure}[ht]
\centering
\includegraphics[width=.97\linewidth]{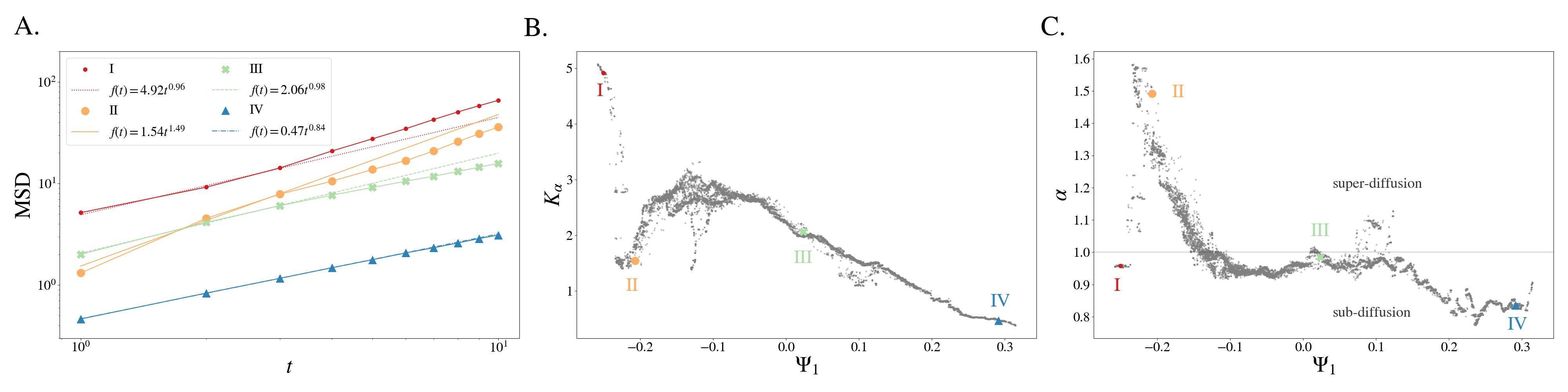}
\caption{(A) MSD for data points within spherical neighborhood of radius $\rho=0.1$ of four selected country-years (see Fig. \ref{fig:diffusion1}). The neighborhood is set in the diffusion space, the MSD is measured in the original 25-dimensional space.
(B) Generalized diffusion coefficient $K_\alpha$ and (C) anomalous diffusion exponent $\alpha$ as a function of DM component $\Psi_1$ (each dot is the measured coefficient for the spherical neighborhood of the corresponding data point).}
\label{fig:diffusion2}
\end{figure}

Next, we computed the MSD in the original 25-dimensional space for each $\rho$-neighborhood in the political diffusion manifold space. Fig. \ref{fig:diffusion2} (A) shows the MSD as a function of $t$ for four example points (the ones represented in Fig. \ref{fig:diffusion1}) in a log-log plot. The almost linear curves obtained for all cases show that the dynamics approximately follows power law over at least a decade. 

Fitting the power law of Eq.~\ref{eqn:diffeq} to each curve yields the generalized diffusion coefficient $K_\alpha$ (Y-intercept) and the anomalous diffusion exponent $\alpha$ (slope) for each neighborhood. 
The fitting is performed over the range $t\in[1,3]$ to capture the dynamics within the boundaries of the considered neighborhood.
Fig. \ref{fig:diffusion2}, which shows $K_\alpha$ (B) and $\alpha$ (C) as a function of $\Psi_1$ for all neighborhoods (i.e. data points).

Our diffusion analysis therefore supports the following conclusions:  

\begin{enumerate}
    \item  Electoral autocracies,exemplified by case I, are typically stable for extended periods of time (sub-diffusive, $\alpha<1$), but when changes occur, they are dramatic and rapid  (high $K_{\alpha}$). An example is Albania during the Cold War (1945--1991), when it was a single-party regime closely aligned with the Soviet Union. The country held elections but did not allow multiparty elections until 1991.
    \item There is a narrow, highly unstable region, exemplified by case II, where states tend to shift quickly (super-diffusive, high $\alpha$) but with smaller steps (low $K_{\alpha}$). These regimes experience more frequent extreme events and shorter stable periods compared to the more autocratic ones. An example is Rwanda from 1978 until 1989, where ethnic violence between Hutus and Tutsis escalated into civil war.
    \item On the broad plateau ($-0.1 \leq \Psi_1 \leq 0.15$) we distinguish two main tendencies: weak sub-diffusion ($\alpha \lesssim 1$) and normal diffusion ($\alpha\sim1$, exemplified by case III).
    An example for both tendencies is Argentina from 1904 until 1983, when the democratisation process started after the military dictatorship known as the National Reorganization Process. 
    In addition, an island of data points shows super-diffusion ($\alpha>1$). They correspond to the restricted democracies located in the upper triangular structure. They leave this region upon a change in suffrage, which is never gradual but from 0.5 to 1 in a single step. %As $\Psi_1$ increases, $K_\alpha$ decreases, meaning that states exhibit change in smaller steps when they have a higher democracy quality.
    \item Full democracies, exemplified by case IV, are the most stable, showing sub-diffusive behavior ($\alpha<1$) with low $K_{\alpha}$, meaning they change little, and when they do, the change tends to be gradual. An example is Norway from 1946 until 2021.
\end{enumerate}

An illustrative example of a country that moves from one dynamic regime into another is Poland, which transitioned from a sub-diffusive state ($\alpha<1$) to a normal-diffusive state ($\alpha \sim 1$) between 2015 and 2019. This shift aligns with findings from other studies, such work based on the Episodes of Regime Transformation (ERT) data, which identified Poland as a backsliding regime since 2015.\cite{ert:2023, ertdata, ertcodebook}

\subsection*{Conflict data}

Given the connection between regime type (as identified here) and movement, plus the relationship between location on the manifold and the risk of regime change, in this section we explore whether location in this space is related to domestic conflict risk. The relationship between armed conflict and political regimes is a key focus in political science that has enormous policy importance but is confounded by democracy measures. Several studies suggest that regimes in the middle range -- between full autocracies and full democracies -- are most prone to war, such that there is an inverted U-shaped relationship between democraticness and conflict risk. \cite{Eck2007,Fein1995, Hegre2001} The empirical support of such studies is relatively mixed or disputed \cite{jones:2018,sambanis:2001}, and empirical studies tend to focus more on regime type than on dynamics. \cite{jones:2018} 
To evaluate whether a country's propensity for conflict is reflected in its movement within the manifold, we use the UCDP/PRIO Armed Conflict Dataset, which provides detailed information on armed conflicts, including location, participants, conflict intensity, and type (such as interstate, intrastate, whether it includes involvement of foreign governments or not), from 1946 to 2022.\cite{Davies2023, Gleditsch2002} 
Here we consider and compare two types of internal armed conflicts covered by the data set, each with a different intensity level: events with between 25 and 999 battle-related deaths in a given year ($\leq 999$ \textit{battle deaths/yr}); and events with at least 1,000 battle-related deaths in a given year ($>999$ \textit{battle deaths/yr}). \cite{Gleditsch2002}

\begin{figure}[ht]
\centering
\includegraphics[width=.95\linewidth]{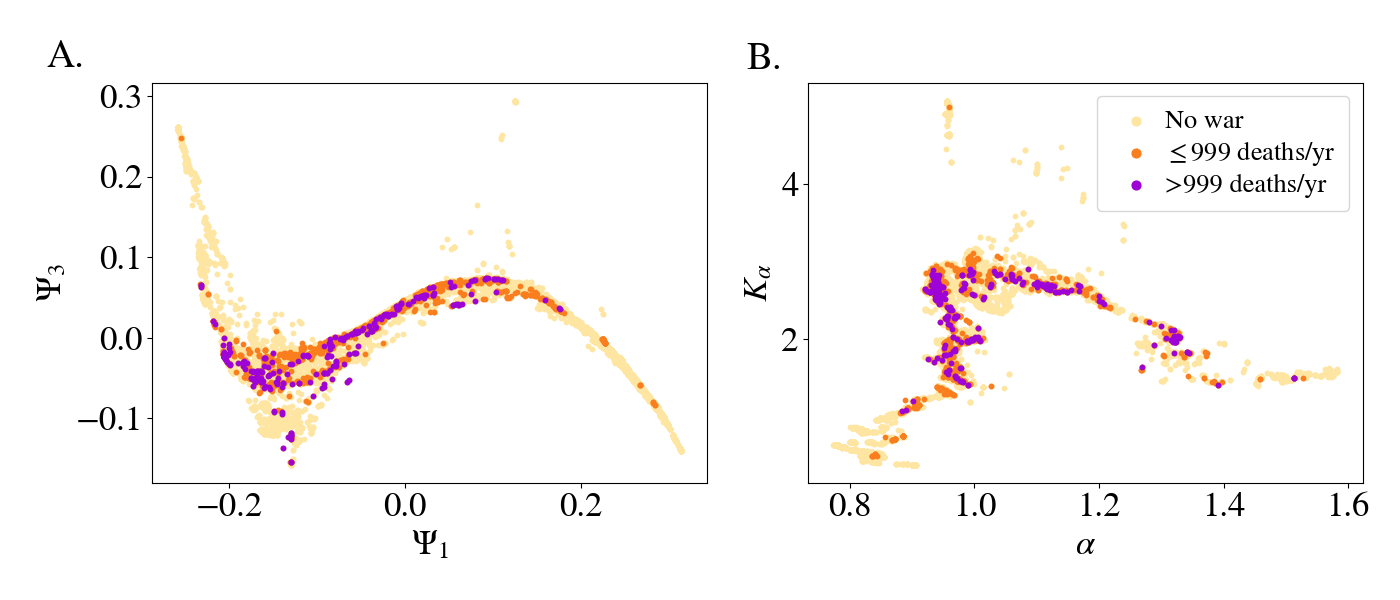}
\caption{Plot of the $\Psi_1$ and $\Psi_3$ projection of the manifold (A) and the anomalous diffusion exponent, $\alpha$, with respect to the generalized diffusion coefficient, $K_{\alpha}$, (B); obtained through the MSD computed taking the neighborhood of each data point. The colors of the markers of all plots indicate the absence (grey) or presence of $\leq 999$ \textit{battle deaths/yr} conflicts (red) and $> 999$ \textit{battle deaths/yr} conflicts (orange) using the UCDP/PRIO Dataset \cite{Davies2023, Gleditsch2002}. The data points (i.e. country-years) plotted are the ones for which both political and conflict data are available.}
\label{fig:conflict}
\end{figure}

In Fig. \ref{fig:conflict}, we visualize the distribution of $\leq 999$ \textit{battle deaths/yr} events and $> 999$ \textit{battle deaths/yr} events in two ways: (A) on the manifold, and (B) as a function of the generalised diffusion coefficient, $K_\alpha$, and the anomalous diffusion exponent, $\alpha$.  
The figure shows that all instances of armed conflict are concentrated in the central region of the manifold, which confirms the hypothesis of the parabolic relationship. \cite{Eck2007,Fein1995, Hegre2001, jones:2018} Additionally, the figure on the right shows that these conflicts primarily occur within specific dynamic regimes, especially in the ``random walk'' area and in the transition area between highly super-diffusive states and random walkers, where $\alpha$ ranges between 0.9 and 1.35. In terms of $K_\alpha$, the specific range for which conflicts occur is between 1.5 and 3. Notably, while not all countries in these regions experience conflict, every country involved in conflict is located there.
In terms of dynamics, countries that are highly super-diffusive or sub-diffusive generally do not experience armed conflicts. In particular, no high-intensity conflicts ($> 999$ \textit{battle deaths/yr}) events are recorded for countries in these regions on the DM. It is not surprising that politically stable regimes do not show armed conflicts, but the absence of such conflicts in highly super-diffusive (i.e. highly unstable) regimes is more surprising. This means that autocratic regimes showing extreme changes are not prone to armed conflict of any intensity level. It is only when these regimes start shifting away from electoral autocracy (in usually fast moves) and begin moving in smaller steps toward the middle of the DM that they show a higher likelihood of armed conflict.

These findings confirm prior research showing that conflict is more likely in the middle ground between full democracies and electoral autocracies. \cite{Eck2007,Fein1995, Hegre2001, jones:2018} Here, however, we add a deeper message: the probability of  conflict is influenced not only by regime type but also by the dynamics of the country. Stable countries are less prone to have armed conflict, but, unexpectedly, highly unstable autocracies also show a lower risk. That offers valuable insight for assessing the likelihood and intensity of future conflicts.

\section*{Discussion and Conclusion} 

In summary, our manifold aligns closely with established indices, such as the Electoral Democracy Index (EDI), yet provides a richer representation by incorporating dynamics. The nonlinearity of the DM technique exposes the varying contribution of  V-Dem variables to  different political regime types, revealing that political regime characterization must go beyond a uniform aggregation of indicators. 
While the manifold construction did not incorporate time, 
it encodes historic trajectories in a meaningful way. 
We have shown that countries evolve in V-Dem space, spanned by election-related variables, as if they were particles that undergo anomalous diffusion. With help of the diffusion-map technique, we were able to relate different regime types to different diffusion processes (see Fig. \ref{fig:diffusion2}). In particular, we showed that democracies correspond to sub-diffusive particles which move slower than normally diffusive particles. 

Anomalous diffusion often stems from specific physical properties of the medium or particle interactions (e.g., traps, obstacles, correlations). 
It is important to point out that in our analysis we did not need to consider any such physical properties to extract the diffusion coefficients from the V-Dem data. 
The fact that different coefficients were found for different parts of the manifold indicates that the diffusion equation provides more than merely a descriptive statistic. 
We also found that countries in armed conflict are mostly random walkers or super-diffusive, linking regime types to conflict dynamics. This finding provides a novel and valuable entryway into assessing conflict risk and intensity.

In addition, our results lend further credibility to the measurement approach of the V-Dem project. The V-Dem data are the result of expert assessments, de-biased as much as that is possible by a Bayesian measurement model. In other words, they are the result of not a physical instrument, with a well-defined error, but of human assessment. The remaining bias in the set-up and conduct of the ‘experiment’ is difficult to estimate. The combination of the V-Dem diffusion manifold with the conflict datasets UCDP/PRIO, however, clearly shows that the manifold constructed from the V-Dem data encodes relevant outcomes, such as the propensity for civil war conflicts.

Nevertheless, our study grossly simplifies a highly complex system.  
For example, we have considered all countries as independent from each other. However, interactions such as trade, geographical proximity, alliances or conflicts affect the political evolution of a country \cite{Coppedge:2022}. It would therefore be interesting in a future study to combine correlation and multi-trajectory analysis to learn about the driving forces (economic, geographic, political, etc).
It is an open question, and subject of ongoing work, whether the anomalous diffusion that we detected can be modeled using a physics-informed approach. The nature of traps, obstacles or long-range correlations, once understood, might give deep insights into the causal mechanisms for regime transformations. 
In this paper we provide the basis for such a physics-informed modeling approach to the V-Dem data and, in general, to  social-science data. 
 Our approach has the potential to provide a foundation for theories of political change, of emergence of conflict, and of extreme events in autocratic (and transitioning) regimes.

\section*{Methods}

\subsubsection*{The V-Dem data set} \label{appendix:data}

We utilize data from the Varieties of Democracy (V-Dem) project, widely regarded as the most detailed source of democracy ratings globally. Developed by the V-Dem Institute in Sweden, this data set evaluates over 200 political units, providing annual ratings from 1789 to the present. V-Dem considers democracy as a multidimensional concept, distinguishing five key principles: electoral, liberal, participatory, deliberative, and egalitarian.  \cite{Coppedge:2022} To quantify these dimensions, the data set offers more than 500 fine-grained indicators for each country-year, which are aggregated into "high-level" indices such as the Electoral Democracy Index (EDI). \cite{vdemmethods}
The data set is built through global collaboration, with around 4,000 country experts answering surveys that assess various democratic attributes, such as media censorship and election fairness. \cite{vdemmethods} A Bayesian Item Response Theory model is applied to correct for systematic biases in the experts' responses and to estimate the uncertainty of the ratings. The raw indicators are combined into lower-level indices like freedom of association and clean elections, which are used to compute the high-level indices. \cite{vdemmethods, vdemcodebook}
Among these high-level indices, the EDI aims to measure Robert Dahl’s concept of \textit{polyarchy}, combining five lower-level indices: freedom of expression, freedom of association, the extent of suffrage, elected officials, and election fairness. \cite{Coppedge:2022, Dahl:1971} In total, 44 indicators are used to calculate the EDI: 9 for freedom of expression, 6 for freedom of association, 8 for clean elections, 20 for elected officials, and 1 for suffrage.

For our analysis, we focus on 25 key variables that contribute to the EDI (see Tab. \ref{tab:var-correl_psi1}). We take all indicators related to freedom of expression, freedom of association and clean elections and share of population with suffrage (\textit{v2x\_suffr}, which is both lower-level index and indicator). For `elected officials' we take the lower-level index (\textit{v2x\_elecoff}), which summarises the 20 indicators related to procedures for electing officials. All indicators in this group are binary (e.g., whether the head of state is directly elected or not) which makes the aggregate index fully informative.

Our data set spans the years 1900 to 2021, offering 12,296 country-year data points. This selection allows for a nuanced analysis of the drivers of democratic processes without the biases introduced by inconsistent variable types.

\begin{table}
\centering
\caption{Name of each variable mentioned in the present study, brief description of each and name of the classification group it belongs to with the V-Dem framework used to construct the EDI \cite{vdemcodebook}.}
\label{tab:var-correl_psi1}
\begin{tabular}{lll}
\textbf{Variable name} & \textbf{Description from \cite{vdemcodebook}} & \textbf{Indicator classification group}                 \\
\hline 
v2x\_suffr             & Share of population with suffrage & Suffrage               \\
v2x\_elecoff           & Elected officials index & Elected officials                         \\
v2psbars               & Barriers to parties & Freedom of association                            \\
v2psparban             & Party ban & Freedom of association                                      \\
v2psoppaut             & Opposition parties autonomy & Freedom of association                    \\
v2cseeorgs             & Civil society entry and exit & Freedom of association                   \\
v2csreprss             & Civil society repression & Freedom of association                       \\
v2elmulpar             & Elections multiparty & Freedom of association                           \\
v2cldiscm              & Freedom of discussion for men & Freedom of expression                  \\
v2cldiscw              & Freedom of discussion for women & Freedom of expression                \\
v2clacfree             & Freedom of academic and cultural expression & Freedom of expression    \\
v2mecenefm             & Government censorship effort – Media & Freedom of expression           \\
v2mecrit               & Print/broadcast media critical & Freedom of expression                 \\
v2merange              & Print/broadcast media perspectives & Freedom of expression             \\
v2meharjrn             & Harassment of journalists & Freedom of expression                      \\
v2meslfcen             & Media self-censorship & Freedom of expression                          \\
v2mebias               & Media bias & Freedom of expression                                     \\
v2elembaut             & Election management body autonomy & Clean elections (I)              \\
v2elintim              & Election government intimidation & Clean elections (I)               \\
v2elfrfair             & Election free and fair & Clean elections (I)  \\
v2elpeace              & Election other electoral violence & Clean elections (II)              \\
v2elembcap             & Election management body capacity & Clean elections (II)              \\
v2elrgstry             & Election voter registry & Clean elections (II)                        \\
v2elvotbuy             & Election vote buying & Clean elections (II)                           \\
v2elirreg              & Election other voting irregularities & Clean elections (II)           
\end{tabular}
\end{table}

\subsection*{The Diffusion Map technique}\label{appendix:dm}

The Diffusion Map (DM) is a nonlinear dimensionality reduction technique based on diffusion processes for finding meaningful and efficient representations of data sets. More precisely, it is a probabilistic interpretation/extension of the spectral embedding dimensionality reduction method. \cite{dm-main, coifman2005geometric, Nadler-dm-chapter, pmlr-vR3-meila01a, NADLER2006113}
It consists of defining a random walk on the data that walks with higher probability to a near data point than to one located far away. 
To describe this process mathematically, pairwise distances between data points are used to define a diffusion operator, i.e. normalized graph Laplacian, which can also be seen as an adjacency matrix of a network where nodes are data points and link weights indicate the proximity between them. Defining the proximity between data points throughout a kernel, the DM acts locally to preserve the lower dimensional structure of the data. Finally, throughout the spectral decomposition of the latter matrix, we can obtain the DM of our data, considering that the eigenvectors with the largest eigenvalues are the ones giving the directions of the largest variation. \cite{dm-cities}

There are several spectral methods based on the spectral decomposition of adjacency matrices, but DMs provide a deeper interpretation. In its definition, Coifman and Lafon define the \textit{diffusion distance} as the distance between two data points based on a random walk in the defined graph. \cite{dm-main} 
Notice that this distance does not necessarily correspond to the Euclidean distance between the same points in the original 25-dimensional space. 
That is exactly the key concept of the DM: the \textit{diffusion distance} in the original space corresponds to the Euclidian distance in the DM space. \cite{dm-main, coifman2005geometric,Nadler-dm-chapter} 
In other words, the \textit{diffusion distance} can be seen as a measure of the connectivity between any pair of data points. Two points can be far away in terms of Euclidean distance, but if they are highly connected via other data points (i.e. there are many data points offering paths in between them), their diffusion distance should be small. 

As described in \cite{dm-main,coifman2005geometric,Nadler-dm-chapter}, the steps to be followed to obtain the DM are:

\begin{enumerate}
    \item Definition of a symmetric and positive semi-definite kernel function. In other words, the probability of walking from data-point $x_i$ to data-point $x_j$ in a single step. A Gaussian kernel with width $\epsilon$ is a common choice: 
    \begin{equation}
        k(x_i,x_j) = exp \left( -\frac{\lVert x_i-x_j \rVert^2}{\epsilon}\right)
    \end{equation}
    Note that for distant $x_i$ and $x_j$, $k(x_i,x_j)\to0$, meaning that trajectories are restricted to the neighbourhood with extension depending on $\epsilon$. This is important in order to drive the diffusion process only through near data-points, capturing the local geometry of the data. Once the kernel is defined, the kernel matrix can be obtained, $K_{ij}=k(x_i,x_j)$.
    \item Construction of the reversible Markov chain, known as the normalized graph Laplacian construction. It represents the probability of transition in one time step from $x_i$ to $x_j$ and it is obtained through the normalisation of the kernel:
    \begin{equation}
        m(x_i,x_j) = \frac{k(x_i,x_j)}{d(x_i)} \quad \text{, where } d(x_i) = \int k(x_i,x_j)d\mu(x_j).
    \end{equation}
    Equivalently, defining $D_{ii}=\sum_j K_{ij}$,
    \begin{equation}
        M = D^{-1}K.
    \end{equation}
    It is worth mentioning that $\int m(x_i,x_j)d\mu(x_j)=1$ and that the probability of transition from $x_i$ to $x_j$ in $t$ time steps is defined by the element $m(x_i,x_j)$ of the transition matrix $M$ to the power of $t$, i.e. $M^t$.
    \item Spectral decomposition of matrix $M$. As it is a stochastic matrix, it can be proven that $M$ has a discrete sequence of left and right eigenvectors, ${\phi_l}$ and ${\psi_l}$, and eigenvalues ${\lambda_l}$ such that $1=\lambda_0 > \lvert \lambda_1 \rvert \geq \lvert \lambda_2 \rvert \geq\dots$ At this point, for computational purposes, one can keep for each data-point only the $\nu$ nearest neighbours just by keeping the first $\nu$ greater values of each row of $M$, setting the rest to zero. \cite{pydiffmap}
    It is interesting to mention the properties of the eigenvectors corresponding to eigenvalue $\lambda_0=1$. 
    On the one hand, since $M$ is a stochastic matrix, then the sum of each row elements is one, which implies $M\overrightarrow{1}= \overrightarrow{1}$. Consequently, the right eigenvector with eigenvalue 1 is an all-ones vector, $\psi_0 = \overrightarrow{1}$. That means that it does not distinguish different nodes of the graph and for this reason it is ignored in the creation of the diffusion map.
    On the other hand, the left eigenvector with eigenvalue 1, $\phi_0$, is, by definition, the stationary distribution of the Markov chain described by $M$. \cite{Nordstrom2005FINITEMC} In addition, $\phi_0(x_i)$ is a density estimate at the point $x_i$. \cite{Nadler-dm-chapter}
    
    % \textit{Proof:} As said in the previous step, running the Markov chain forward in time corresponds to compute $M^t$, being $t$ the number o time steps we want to consider. The matrix $M^t$ can be written in terms of its left and right eigenvectors, ${\phi_l}$ and ${\psi_l}$, and eigenvalues ${\lambda_l}$, as:

    % \begin{align}
    % M^t &=
    % \begin{bmatrix}
%     \psi_0 \dots \psi_n
%     \end{bmatrix}
%     \begin{bmatrix}
% \lambda_0 & 0   & \dots & 0 \\
% 0   & \lambda_1 & \dots & 0 \\
% \vdots & \vdots & \ddots & \vdots \\
% 0   & 0   & \dots & \lambda_n \\
% \end{bmatrix}^t
%     \begin{bmatrix}
%     \phi_0  \\ \vdots \\ \phi_n
%     \end{bmatrix} \\
%     &= 
%     \begin{bmatrix}
%     \psi_0 \dots \psi_n
%     \end{bmatrix}
%     \begin{bmatrix}
% \lambda_0^t & 0   & \dots & 0 \\
% 0   & \lambda_1^t & \dots & 0 \\
% \vdots & \vdots & \ddots & \vdots \\
% 0   & 0   & \dots & \lambda_n^t \\
% \end{bmatrix}
%     \begin{bmatrix}
%     \phi_0  \\ \vdots \\ \phi_n
%     \end{bmatrix} \\
%     &= \lambda_0^t\psi_0\phi_0+\lambda_1^t\psi_1\phi_1+\dots+\lambda_n^t\psi_n\phi_n
%     \end{align}
    
%     Then, as $1=\lambda_0 > \lvert \lambda_1 \rvert \geq \lvert \lambda_2 \rvert \geq\dots\geq \lvert \lambda_n \rvert$, 
%     \begin{equation}
%         \lim_{t\to\infty} M^t = \psi_0\phi_0 = \phi_0 \hspace{.3cm} \blacksquare
%     \end{equation}
    \item Definition of the family of diffusion maps $\{\Psi_{t}\}_{t\in\mathbb{N}}$,
    \begin{equation}
        \Psi_t(x) = \begin{pmatrix}
                    \lambda_1^t \psi_1(x)\\
                    \lambda_2^t \psi_2(x)\\
                    \vdots \\
                    \lambda_{s(\delta, t)}^t \psi_{s(\delta, t)}(x)
                    \end{pmatrix}.
    \end{equation}
    Each $\Psi_t(x)$ component is a diffusion component or coordinate and the whole map embeds the original data into a new space of $s(\delta, t)$ dimensions, being $\delta$ the relative accuracy. 
    In addition, the components corresponding to the largest eigenvalues (i.e. the first ones), correspond to the directions of slower diffusion. As shown in Eqs. 6 and 7, the larger the corresponding eigenvalue, the slower the convergence of the diffusive process to the stationary state.
    In the end, the DM $\Psi_t(x)$ embeds the data into a new Euclidean space $\mathbb{R}^{s(\delta,t)}$, i.e. the DM space, where the Euclidean distance is the \textit{diffusion distance} when $s=n-1$:
    \begin{equation}
        D_t^2(x_0, x_1)=\sum_{j\geq1} \lambda_j^{2t}(\psi_j(x_0)-\psi_j(x_1))^2=\lVert \Psi_t(x_0)-\Psi_t(x_1) \rVert^2
    \end{equation}
\end{enumerate}

Once the DM coordinates, $\Psi_{t}$, are obtained, their relationships must be analyzed. As said in \cite{Nadler-dm-chapter}, several eigenvectors can be redundant by encoding the same geometrical or spatial ``direction" of a manifold. For this reason, relations between the different components must be studied to remove this redundancy and find the most sensible representation of the data.

The DM technique has been applied to our data defining $(\epsilon, t, \nu)=(10,1,100)$. These values have been chosen after testing $\epsilon \in (1,10000)$ and $\nu \in$ (50, \text{all}), and seeing that no significant difference is appreciated in the structures obtained for a quite wide range around the values chosen.

From all coordinates, the three first ones have been selected for several reasons. On the one hand, the spectral decomposition of $M$ in the DM creation leads to a sequence of eigenvalues, decreasing both in value and relevance. \cite{dm-main} As shown before, the Markov chain in the DM algorithm identifies fast and slow directions of propagation. \cite{dm-main} This implies that the diffusion time scale along each component $i$ is inversely proportional to its corresponding eigenvalue $\lambda_i$. In physical terms, eigenvectors with larger eigenvalues indicate the directions of the slowest diffusion, and these directions encapsulate the most significant geometric information of the data. Again, the first DM coordinates are the ones corresponding to larger eigenvalues.  On the other hand, the manifold projected onto dimensions $\Psi_1$, $\Psi_2$ and $\Psi_3$ encapsulates the history of the 20th century. As shown and explained in the main text, we can interpret the structure in terms of historical events, which supports the significance of the newly defined space.

\subsection*{Three-dimensional fitting of the political diffusion manifold}\label{appendix:fit}

We fitted the three-dimensional manifold with the function \textit{curve\_fit} of the Python library SciPy \cite{2020SciPy}, which from the next parameterized curve,

\begin{equation}
    \begin{cases}
        x = \Psi_1 \\
        y = ax^3 + bx^2 + cx + d \\
        z = ex^5 + fx^4 + gx^3 + hx^2 + ix + j
    \end{cases}
\end{equation}

led to the following fitting parameters:

\begin{equation}
    \begin{aligned}[c]
        \begin{cases}
            a = -6.8620 \\
            b = 4.6276\\
            c = 0.0484\\
            d = -0.1082
        \end{cases}
    \end{aligned}
\qquad\qquad
    \begin{aligned}[c]
        \begin{cases}
        e = -78.0610 \\
        f = 70.6115\\
        g = -17.1926\\
        h = -3.8002\\
        i = 0.8833\\
        j =0.04868 .
        \end{cases}
    \end{aligned}
\end{equation}
We projected all data points onto this curve and obtained a simplified, unfolded version of our DM manifold. To do so, we considered the data point being at one of both extremes of the manifold, $X_0$, to be at the origin of the new coordinate $s$. In other words, $s(X_0)=0$. From there, we identified the nearest neighbour, $X_1$, and computed the Euclidean distance between them, $d_{0,1}$, which defines the position of data point $X_1$ in the new coordinate system as $s(X_1)=d_{0,1}$. We repeated the same process with the nearest neighbour of $X_1$, which is $X_2$, and computed their Euclidean distance, $d_{1,2}$. And this time, $s(X_2)=d_{0,1}+d_{1,2}$. This method can be summarized by the equation $s(X_i) = \sum_{j = 0}^{i-1}d_{j,j+1}$, being $d_{j,j+1}$ the Euclidean distance between points $X_j$ and $X_{j+1}$. Following this process for the whole set of data points, we end up with a one dimensional coordinate, here referred to as $s$, that encapsulates the manifold while preserving both the order and distances between neighboring points in the projection. 

\bibliographystyle{unsrt}
\bibliography{sample}

\section*{Acknowledgements}

We thank Andrey G. Cherstvy, Philipp Meyer and Maximilian Graf for useful discussions.

\section*{Author contributions statement}

K.W. and M.C.W. conceptualised and supervised the study, P.P.D. and S.B. wrote the code, P.P.D. analyzed the data, all authors interpreted the results, P.P.D. wrote the initial draft, all authors edited the draft.

\section*{Additional information}

The authors declare that they have no known competing financial interests or personal relationships that could have appeared to influence the work reported in this paper.

\end{document}